\documentclass[proof]{pasj00}
\draft

\SetRunningHead{Aoki et al.}{SDSS~1723+5553}
\Received{2010/3/22}
\Accepted{2010/07/26}
\Published{}

\begin{document}

\title{Broad Balmer-line Absorption in SDSS~J172341.10+555340.5
\thanks{Based on data collected at Subaru Telescope, which is operated by the National Astronomical Observatory of Japan.}}

\author{Kentaro \textsc{Aoki}}
\affil{Subaru Telescope, National Astronomical Observatory of Japan,
    650 North A'ohoku Place, Hilo, \\HI 96720, U.S.A.}

\maketitle

\begin{abstract}
I present the discovery of Balmer-line absorption from H$\alpha$ to H9 in iron low-ionizaton broad absorption line (FeLoBAL) quasar, 
SDSS~J172341.10+555340.5 by near-infrared spectroscopy
with the Cooled Infrared Spectrograph and Camera for OHS (CISCO)
attached to the Subaru telescope.
The redshift of the Balmer-line absorption troughs is $2.0530\pm0.0003$,
and it is blueshifted by 5370 km s$^{-1}$ from the
Balmer emission lines.
It is more than $4000$ km s$^{-1}$
blueshifted from the previously known UV absorption lines.
I detect relatively strong (EW$_{\rm rest}=20$ \AA) [O \emissiontype{III}] emission lines 
which are similar to those found in other broad absorption line quasars with 
Balmer-line absorption. 
I derived a column density of neutral hydrogen of $5.2\times10^{17}$ cm$^{-2}$
by using the curve of growth and taking account of Ly$\alpha$ trapping. 
I searched for UV absorption lines which have the same redshift with Balmer-line absorption.
I found Al \emissiontype{III} and Fe \emissiontype{III} absorption lines at $z$=2.053 which correspond to previously unidentified absorption lines, and the presence of other blended troughs that were difficult to identify.
\end{abstract}

\section{Introduction}
Outflow phenomena are ubiquitous among active galactic nuclei (AGN)
in several regimes of the full electromagnetic spectrum.
Recent X-ray spectroscopic observations reveal blueshifted absorption lines from highly ionized species in many nearby Seyfert 1s (see reviews by \citet{TM09}).
Since their early discovery by \citet{Lynds1967}, 
broad absorption lines (BALs) in quasars have been well studied.
The widths of BAL are typically $\sim2000-20,000$ km s$^{-1}$,
and are blueshifted relative to emission lines from 2000 km s$^{-1}$
to as much as $\sim 0.2c$.
Half of all quasars have intrinsic narrow absorption lines that have FWHMs of a few hundred km s$^{-1}$
and most often blueshifted from the emission-line redshift in quasars \citep{Misawa07, GB08}.
Similar absorption lines are common in nearby Seyfert 1 galaxies.
More than half of  Seyfert 1s show C \emissiontype{IV} and N \emissiontype{V} absorption lines in their
UV spectra \citep{Cre1999}, and half of them show O \emissiontype{VI} absorption \citep{Dunn07}.
\citet{CK05} point out that inner narrow-line region (NLR) clouds
are the source of UV absorbers in the Seyfert galaxy NGC~4151
based on their similar kinematical characteristics and locations.
\citet{Cec02} also suggests that the high-velocity clouds in NLR of Seyfert 2 galaxy NGC~1068
would resemble intrinsic narrow absorption lines if they were viewed against the nuclear continuum source.
\par
It is important to understand outflows because they transfer
nuclear gas from near the super massive black hole (SMBH) to the outer region of host galaxies.
Thus, these outflows may be related to the co-evolution of the bulge and SMBH \citep{SR98, Fab99, Gra04}.
We do not yet fully understand not only the theory of outflow phenomena
but also the phenomenologies.
Recent large surveys such as the Sloan Digital Sky Survey (SDSS; \cite{Yor00}) have discovered outflows from ions not previously detected \citep{Hal02}.
One recent example is Balmer-line BALs \citep{Aok06, Hal07}. \par
We discovered H$\alpha$ absorption in the broad H$\alpha$ emission line of the 
BAL quasar, SDSS~J083942.11+380526.3 (hereafter SDSS~J0839+3805) through near-infrared 
spectroscopy \citep{Aok06}. 
The presence of nonstellar H$\alpha$ absorption was known only in the Seyfert galaxy NGC~4151 at that time; 
thus, our discovery was the first case for quasars.
SDSS~J0839+3805 is a so-called many-narrow-trough FeLoBAL quasar (hereafter mntBAL) \citep{Hal02},
which has tremendous number of UV absorption lines.
\citet{Hal07} also discovered Balmer absorption lines in SDSS~J125942.8+121312.6 (hereafter
SDSS~J1259+1213) from its SDSS spectrum.
SDSS~1259+1213 is also a probably mntBAL.
Motivated by these discoveries, I began a search for Balmer-line absorption 
in mntBALs.
SDSS~J172341.10+555340.5 (hereafter SDSS~J1723+5553) is a mntBAL reported by \citet{Hal02}, and it is a bright quasar
in the near-infrared ($J=16.1$, and $K_{s}=14.1$).
I performed near-infrared spectroscopy of SDSS~J1723+5553,
and report in this paper my discovery of the Balmer line absorption in SDSS~J1723+5553.
I assume throughout this paper $H_{\rm 0} = 70$ km s$^{-1}$ Mpc$^{-1}$, 
$\Omega_{m} = 0.3$, and $\Omega_{\Lambda} = 0.7$.
Note that all wavelengths in this paper are vacuum wavelengths.
\par

\section{Observations and Data Reduction} \label{Obs}
The $JH$- and $K$-band spectra of SDSS~J1723+5553 were obtained 
with the Cooled Infrared Spectrograph and Camera for OHS 
(CISCO; \cite{Mot02})
attached to the Subaru 8.2 m telescope \citep{Iye04} on 2005 July 16 (UT), 
which was relatively clear night with good seeing (\timeform{0.5''} in $K$ band). 
I set the slit width to \timeform{0.6''}, which 
results in a resolution of 56 and 60 {\AA}  in $JH$- and $K$-band 
(i.e., $R\sim 300 - 350$), respectively, which were measured by using night sky 
lines. 
The slit position angle was \timeform{0D}. 
I obtained four and two exposures in $JH$- and $K$-band, respectively,
dithering the telescope to observe the quasar at two positions with a 
separation of \timeform{10''} along the slit. 
The integration times on target were 1200 s (300 s $\times 4$) in $JH$ band and 500 s (250 s $\times 2$) 
in $K$ band. 
The A3 star SAO~030517 was observed immediately after exposures of 
SDSS~J1723+5553 for sensitivity calibration and removal of atmospheric absorption lines.

The data were reduced using IRAF for the standard procedures of flat-fielding,
sky subtraction, and residual sky subtraction. 
Wavelength calibration was performed using OH night-sky lines. 
The rms wavelength calibration error is 1.5 {\AA} in $JH$- and 6.0 {\AA} in $K$-
band, 
corresponding to 36 km s$^{-1}$ at 16170 {\AA} (redshifted position of the H$\beta$ line of SDSS~J1723+5553) 
and 80 km s$^{-1}$ at 21780 {\AA}  (redshifted H$\alpha$ line), respectively. 
The sensitivity calibration was performed as a function of wavelength, 
and the atmospheric absorption features were removed with the spectrum of SAO~030517.

\section{Results} \label{Results}
\begin{figure}
  \begin{center}
    \FigureFile(135mm,110mm){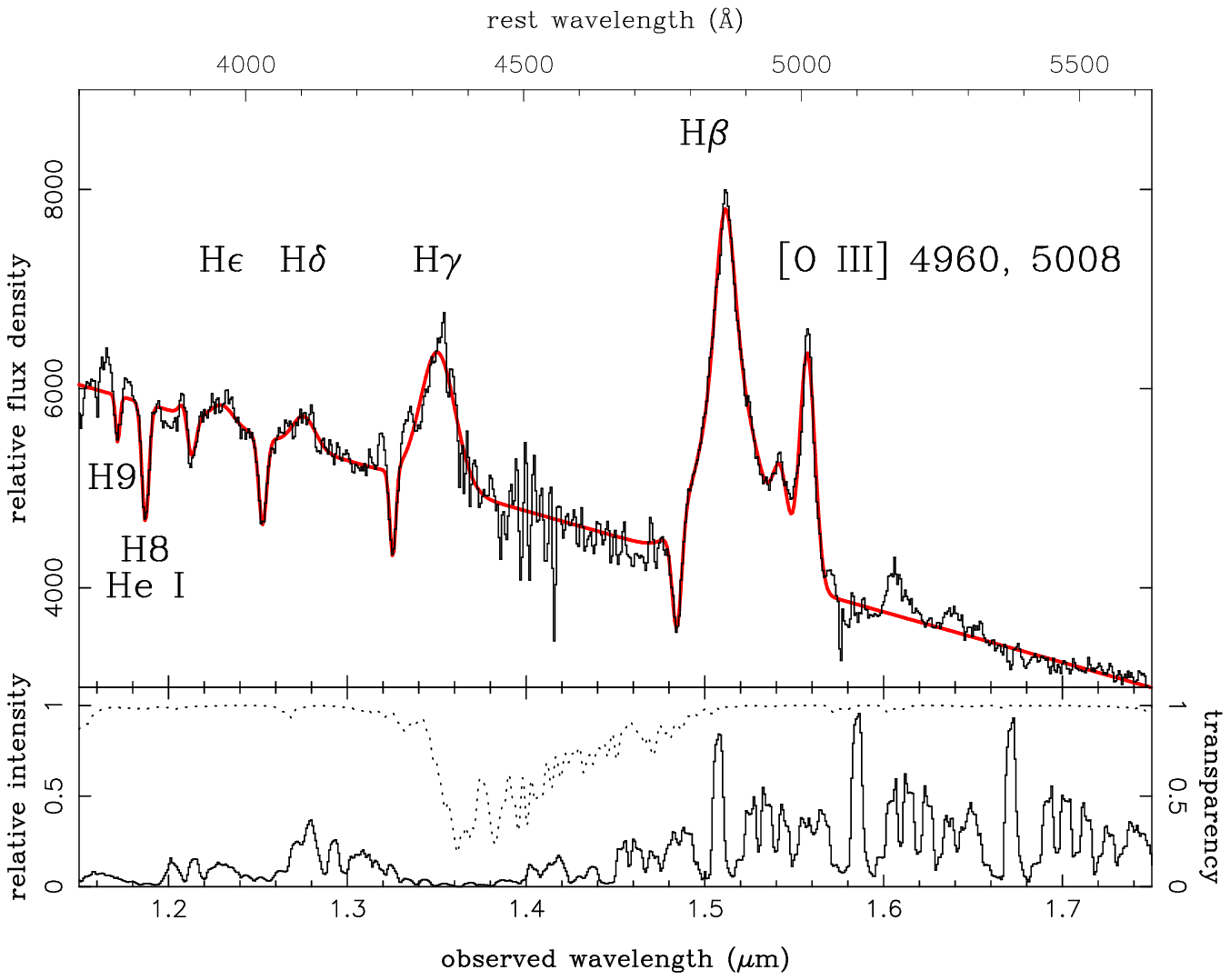}
  \end{center}
  \caption{Top: $JH$-band spectrum of SDSS~J1723+5553.
Ordinate is a relative flux density in units of erg s$^{-1}$ cm$^{-2}$ $\micron^{-1}$, 
and abscissa is the observed wavelength in vacuum in microns.
The rest wavelength is given along the top axis.
The best fit result is shown as a red line.
Bottom: The sky emission spectrum (solid line) and the atmospheric transmission curve 
(dotted line), which is obtained from the United Kingdom Infra-Red Telescope (UKIRT)
Web page. It is produced using the program IRTRANS4.}
\label{fig1}
\end{figure}

Figure \ref{fig1} displays the $JH$-band spectrum of 
SDSS~J1723+5553.
The $JH$-band spectrum clearly shows several absorption lines near the
H$\beta$ emission line and [O \emissiontype{III}] emission lines.
H$\gamma$ emission line is also identified although it overlaps with 
the atmospheric absorption lines.
The small bumps at 12760 {\AA}  and 12300 {\AA}  correspond to H$\delta$ and H$\epsilon$
emission lines, respectively.
The absorption lines must be Balmer-line absorption since most of them occur
at the edge of blue wing of the broad Balmer emission lines.
I fit the $JH$-band spectrum with a combination of a linear continuum,
four Balmer emission lines from H$\beta$ to H$\epsilon$, 
six Balmer absorption lines from H$\beta$ to H9, 
and [O \emissiontype{III}] $\lambda\lambda 4960, 5008.2$ emission lines.
All Balmer emission and absorption lines are fitted with a single Gaussian except for the H$\beta$ emission line, which is
fitted with a combination of two Gaussians.
I fit the [O \emissiontype{III}] doublet $\lambda\lambda 4960.3, 5008.2$ with a single Gaussian for each line. 
The width and redshift are assumed to be the same for both of the two [O \emissiontype{III}] lines 
and an intensity ratio of [O \emissiontype{III}] $\lambda5008/\lambda4960$ 
fixed to be 3.0.
The FWHM is corrected for the instrumental broadening by using the simple
assumption:
$\rm{FWHM}_{true}=(\rm{FWHM}_{obs}^{2} - \rm{FWHM}_{inst}^{2})^{1/2}$,
where $\rm{FWHM}_{obs}$ is the observed FWHM of the line and
$\rm{FWHM}_{inst}$ is the instrumental FWHM.
The results of fitting are tabulated in table \ref{tbl1}.
The $\rm{FWHM}_{true}$ of H$\beta$ emission line is $3780\pm150$ km s$^{-1}$.
The redshifts of Balmer emission lines from H$\beta$ to H$\epsilon$ vary from 2.097 to 2.111.
However, the H$\beta$ emission line has much better signal-to-ratio than other Balmer emission lines.
Also, the redshift of the H$\alpha$ emission line described below is same as that of the H$\beta$ emission line.
Therefore, I adopt $2.1080\pm0.0001$ as the redshift of broad emission line.
\par
The redshifts of the Balmer-line series absorption troughs range from 2.0527 (H$\beta$) to 2.0540 (H9).
The redshift of the H8 ($\lambda 3890.2$\AA) absorption line is smaller than other Balmer lines 
due to probably contamination of He \emissiontype{I} 3889.7. 
Since the H9 absorption line is much weaker than the other Balmer absorption lines,
I adopt $2.0530\pm0.0003$ as the redshift of the system.
This blueshift translates to an outflow velocity of 5370 km s$^{-1}$
and 
a $\rm{FWHM}_{obs}$ of $700 -1500$ km s$^{-1}$.
Because the $\rm{FWHM}_{inst}$ is 1160 km s$^{-1}$ in the $JH$ band,
the lines are not fully resolved.
Although the $\rm{FWHM}_{true}$ of the absorption lines have large uncertainties,
the FWHM of the Balmer lines are consistent with Fe~\emissiontype{III} and Al~\emissiontype{III} absorption lines at the same redshift (described in \ref{Metal}).
I present the equivalent widths (EWs) of the Balmer absorption troughs in table \ref{tbl2}.
I do not include broad emission lines for the continuum levels when I 
calculate the EWs.
\par 
I detect the relatively strong (EW$_{rest}=20.3\pm0.1$\AA)
[O \emissiontype{III}] emission lines, which 
is common among four Balmer-lines BAL AGN (NGC~4151, SDSS~0839+3805, 
SDSS~J1259+1213 and SDSS~J1723+5553).
I find a redshift for the [O \emissiontype{III}] emission of $2.1095\pm0.0004$.
Thus, the [O \emissiontype{III}] emission line is redshifted by 145 km s$^{-1}$
relative to the broad emission line.
The width of the [O \emissiontype{III}] emission line is resolved and 
the $\rm{FWHM}_{true}$ is $1550\pm70$ km s$^{-1}$.
The optical luminosity of SDSS~1723+5553, $\log \lambda L_{\lambda} (5100)$
is 47.02 erg s$^{-1}$ after correction for reddening ($E(B-V)=0.25$ mag).
Compared with similar luminous quasars \citep{Net04},
this line width is normal.
\par
\begin{figure}
  \begin{center}
    \FigureFile(135mm,110mm){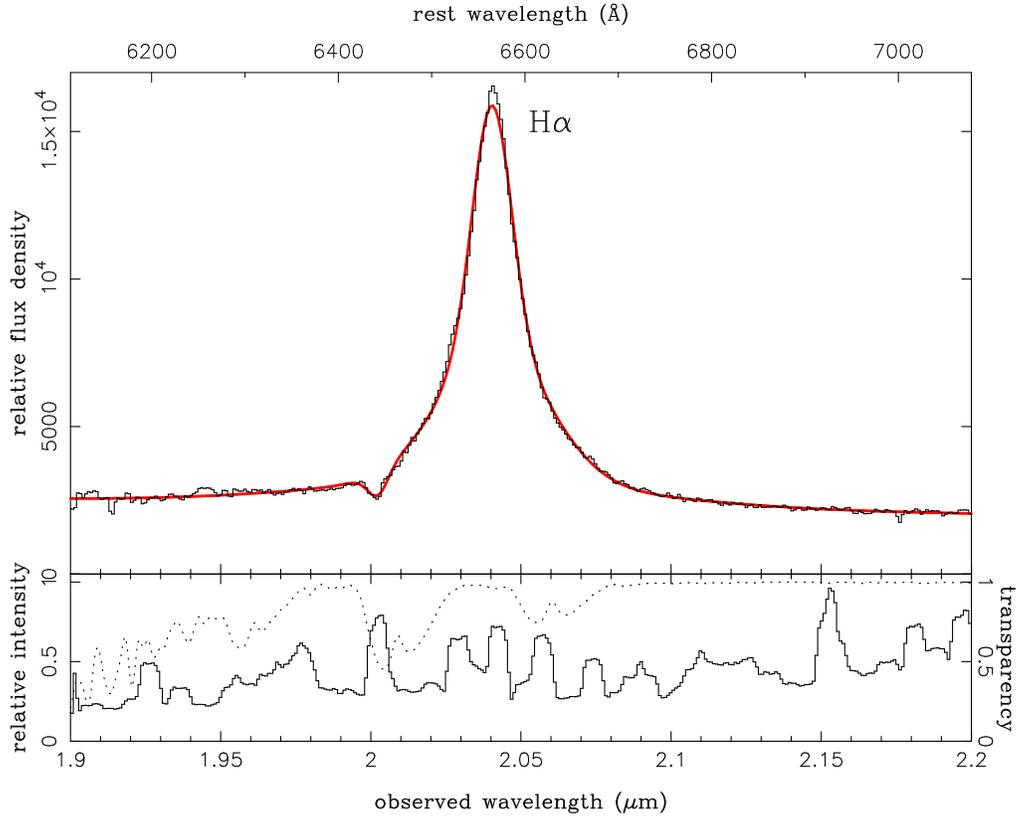}
  \end{center}
  \caption{Top: $K$-band spectrum of SDSS~J1723+5553.
Ordinate is a relative flux density in units of erg s$^{-1}$ cm$^{-2}$ $\micron^{-1}$, 
and abscissa is the observed wavelength in vacuum in microns.
The rest wavelength is given along the top axis.
The H$\alpha$ emission line is fitted with three Gaussians.
The best fit is shown as a red solid line.
Bottom panel is as in Fig. 1.}
\label{fig2}
\end{figure}

The $K$-band spectrum is displayed in Fig. \ref{fig2}.
The strong H$\alpha$ emission is associated with a weak absorption line.
I fit the $K$-band spectrum with a linear continuum, H$\alpha$ broad emission line, and H$\alpha$ absorption line.
The H$\alpha$ emission line is  modeled with a combination of three Gaussians,
and the H$\alpha$ absorption line is fitted with a single Gaussian.
The results are tabulated in table \ref{tbl1} and \ref{tbl2}.
The redshift of the H$\alpha$ emission line agrees with that of the H$\beta$ emission line.
The $\rm{FWHM}_{true}$ of H$\alpha$ is 2780 km s$^{-1}$ and
it is 40\% smaller than the $\rm{ FWHM}_{true}$ of H$\beta$.
This difference between the width of H$\alpha$ and H$\beta$ is larger than 
the 17\% found by \citet{GH05} among 160 sample of AGN.
\par 
The redshift of the H$\alpha$ absorption line is smaller than that of the H$\beta$ absorption line.  
However, the H$\alpha$ absorption line is overlapped with the strong atmospheric absorption line at $2.00\micron$ (figure \ref{fig2}).
Thus, the redshift of the H$\alpha$ absorption line is less accurate than that of H$\beta$ absorption line.

\section{Discussion} \label{Discussion}
\subsection{Balmer absorption lines}
The Balmer absorption lines found in SDSS~J1723+5553 are from an outflow phenomenon
from the nucleus, not from the host galaxy.
First, the Balmer absorption lines are blueshifted by 5370 km s$^{-1}$ from
the emission lines.
Second, there are Fe \emissiontype{III} and Al \emissiontype{III} absorption lines at the same redshift of the Balmer absorption lines (see \S 4.2).
This is not the case in post-starburst galaxies which exhibit strong Balmer absorption lines.       
\par
\begin{figure}
  \begin{center}
    \FigureFile(135mm,110mm){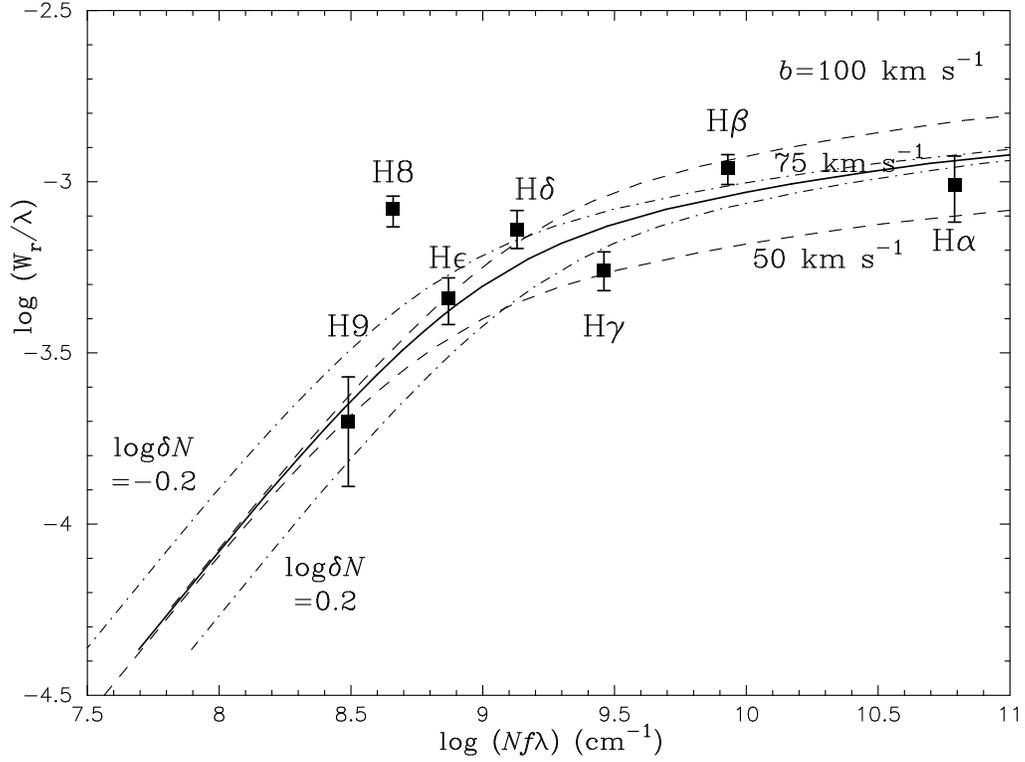}
  \end{center}
  \caption{The growth curve of Balmer-lines absorption.
The best-fit curve of growth curve (75 km s$^{-1}$ of $b$ and $1.5 \times 10^{15}$ cm$^{-2}$ of the column density) is indicated by the solid line.
The dashed-lines indicate the cases of 25 km s$^{-1}$ larger and smaller than the best Doppler parameter.
The do-dashed lines indicate the cases of 0.2 larger and smaller than the best column density in logarithmic scale.
The strength of H8 clearly deviates from the curve of growth.
This supports possible contamination of He I.
I did not use H8 for the fitting.}\label{fig3}
\end{figure}

In order to derive the column density of neutral hydrogen from the unresolved Balmer absorption lines,
I use the curve of growth \citep{Sp78}.
I fit for the Doppler parameter, $b$,
and the column density of neutral hydrogen at the n=2 level (e.g., \cite{Sav03}).
The resulting best-fit curve of growth is shown in figure \ref{fig3}.
I derived $(1.5 \pm 0.9) \times 10^{15}$ cm$^{-2}$ for the hydrogen column density at 
n=2 level and $75 \pm 25$ km s$^{-1}$ for $b$.  
As pointed out in \S 3, the H8 absorption line is probably contaminated by a He I absorption line.
The strength of H8 clearly deviates from the curve of growth in figure \ref{fig3}.
This also supports possible contamination of He I, and therefore
I do not use H8 for the fitting.
The Doppler parameter $b=75$ km s$^{-1}$ corresponds to a FWHM of 125 km s$^{-1}$.
This FWHM is much smaller than those I measure in \S3 (700 - 1500 km s$^{-1}$).
This fact suggests that the absorption lines consist of separated
narrow (FWHM$=100-200$ km s$^{-1}$), but not strongly saturated components. 
Such narrow separated components are in fact observed with high resolution spectroscopy in FeLoBALs, QSO 2359-1241 \citep{Arav08},
SDSS J0318-0600 \citep{Dunn10} and AKARI-IRC 1757+5907 
(Aoki et al, in preparation).
In such cases, velocity width derived from low resolution spectroscopy indicates
a relative distribution in velocity range of narrow components, not the width of each component.
However, the equivalent width is not affected by the low resolution.
The Doppler parameter derived from the curve of growth method should be much smaller than the velocity width measured by low resolution spectroscopy.
Since I do not resolve the absorption lines, I can not estimate the covering factor.
Thus, I assume the covering factor of the absorber of the continuum is 1.
If the covering factor is significantly smaller than 1, 
as estimated for SDSS~J1259+1213 by \citet{Hal07}, 
the column density estimated above would be lower limit.
\par
The optical depth at the center of the H$\beta$ absorption line ($\tau_{H\beta}$) 
is estimated to be 17
(Eq. (3-51) of \citet{Sp78}).
Such a large $\tau_{H\beta}$ results in a large $\tau_{Ly\alpha}$
since $\tau_{H\beta}$ and $\tau_{Ly\alpha}$ follow the relationship,
\begin{equation}
\tau_{Ly\alpha}=\frac{\lambda_{Ly\alpha} f_{Ly\alpha}}{\lambda_{H\beta}
f_{H\beta}} \frac{N_{1}}{N_{2}} \tau_{H\beta} = 
0.87  \frac{N_{1}}{N_{2}} \tau_{H\beta},
\end{equation}
where $\lambda$ is wavelength, $f$ is oscillator strength, and $N_{1}$ and 
$N_{2}$ are the population of the level n=1 and n=2, respectively  
\citep{Hal07}.
As pointed out by \citet{Hal07},
when $\tau_{Ly\alpha}$ is large,
every Ly$\alpha$ photon created by recombination is absorbed $\tau_{Ly\alpha}$ 
times before it escapes, and excites a hydrogen atom to its n=2 level.
The n=2 level population for a given temperature is thus increased by a factor of 
$\tau_{Ly\alpha}$
from the thermal equilibrium,
\begin{equation}
\frac{N_{1}}{N_{2}} = \frac{1}{4} \exp ({\rm 10.2 eV} / kT) \frac{1}{\tau_{Ly\alpha}} 
\end{equation}
\citep{Hal07}.
Substituting our $\tau_{H\beta}$ and $T=7500$ K \citep{OF06} into (1) and (2), 
I derive $\tau_{Ly\alpha} = 5140$, and 
$\frac{N_{1}}{N_{2}} = 348$.
The neutral hydrogen column density is
derived to be $5.2 \times 10^{17} {\rm cm}^{-2}$.
\par
\subsection{Metal absorption lines at z=2.053}\label{Metal}
\begin{figure}
  \begin{center}
    \FigureFile(135mm,110mm){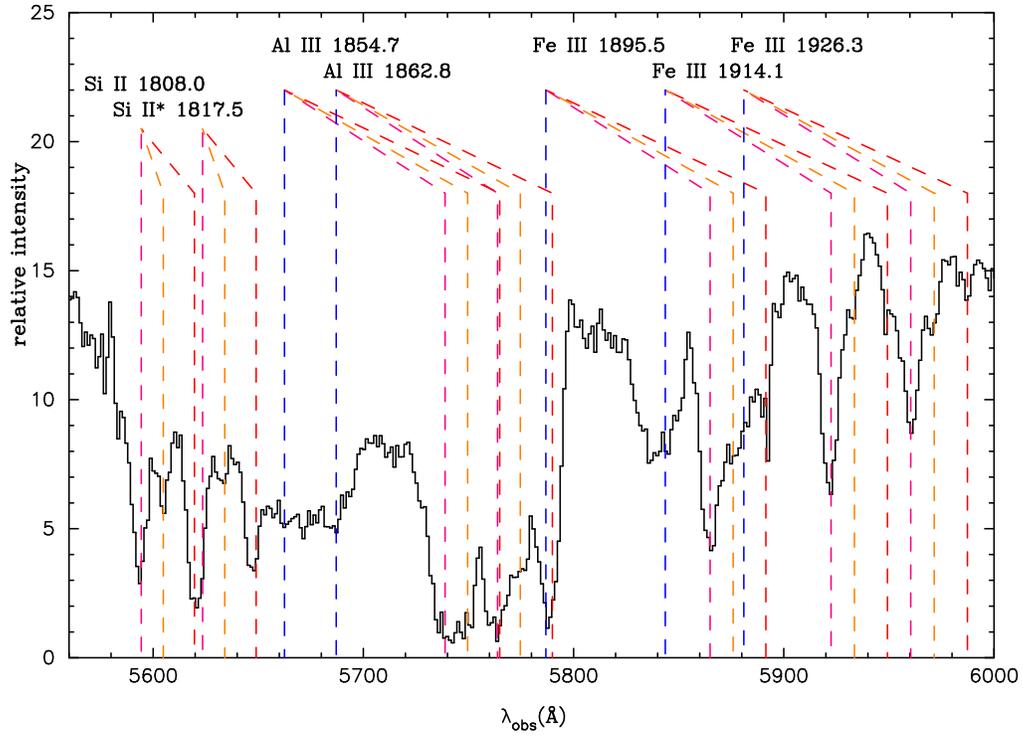}
  \end{center}
  \caption{Search for $z=2.053$ system.
The spectrum of SDSS~J1723+5553 between 5540 {\AA} and 6000 {\AA} 
in observed frame is shown.
The known absorption systems at $z=2.0942, 2.100$ and 2.1082 are indicated by
magenta, orange and red dashed lines, respectively.
The expected positions of Al \emissiontype{III} and Fe \emissiontype{III} 
absorption lines at $z=2.053$
are indicated by blue dashed lines.
}\label{fig4}
\end{figure}

The Balmer-line absorption of SDSS~J1723+5553 I have discovered has a redshift of 2.053.
It is $> 4000$ km s$^{-1}$
blueshifted from the previously known absorption lines.
\citet{Hal02} identified three absorption lines systems ($z=2.0942, 2.100,$ and $2.1082$) in the rest UV spectrum.
In order to check whether $z=2.053$ UV metal absorption lines exist,
I examine the UV spectrum of SDSS~J1723+55553 from the SDSS Data Release 5.
I have searched for the $z=2.053$ absorption lines from the list by \citet{Hal02} between
3800 {\AA} and 9000 {\AA}, which corresponds to a rest wavelength range of 1245 {\AA} and 2947 {\AA}.
Many absorption lines from the $z=2.0942, 2.100$ and 2.1082 systems exist in the UV spectrum.
Because these troughs are blending,
it is difficult to identify $z=2.053$ absorption lines.
At least Fe \emissiontype{III} UV34 $\lambda\lambda1914.1, 1926.3$ and Al \emissiontype{III} $\lambda\lambda1854.7, 1862.8$ 
at $z=2.053$ are convincingly identified (figure \ref{fig4})
although other absorption at $z=2.053$ probably exists and
is hidden by severe overlap with absorption at $z=2.0942, 2.100$ and $2.1082$.
I describe in detail
the identifications of those three absorption in following paragraphs.
\par
\begin{figure}
  \begin{center}
    \FigureFile(170mm,140mm){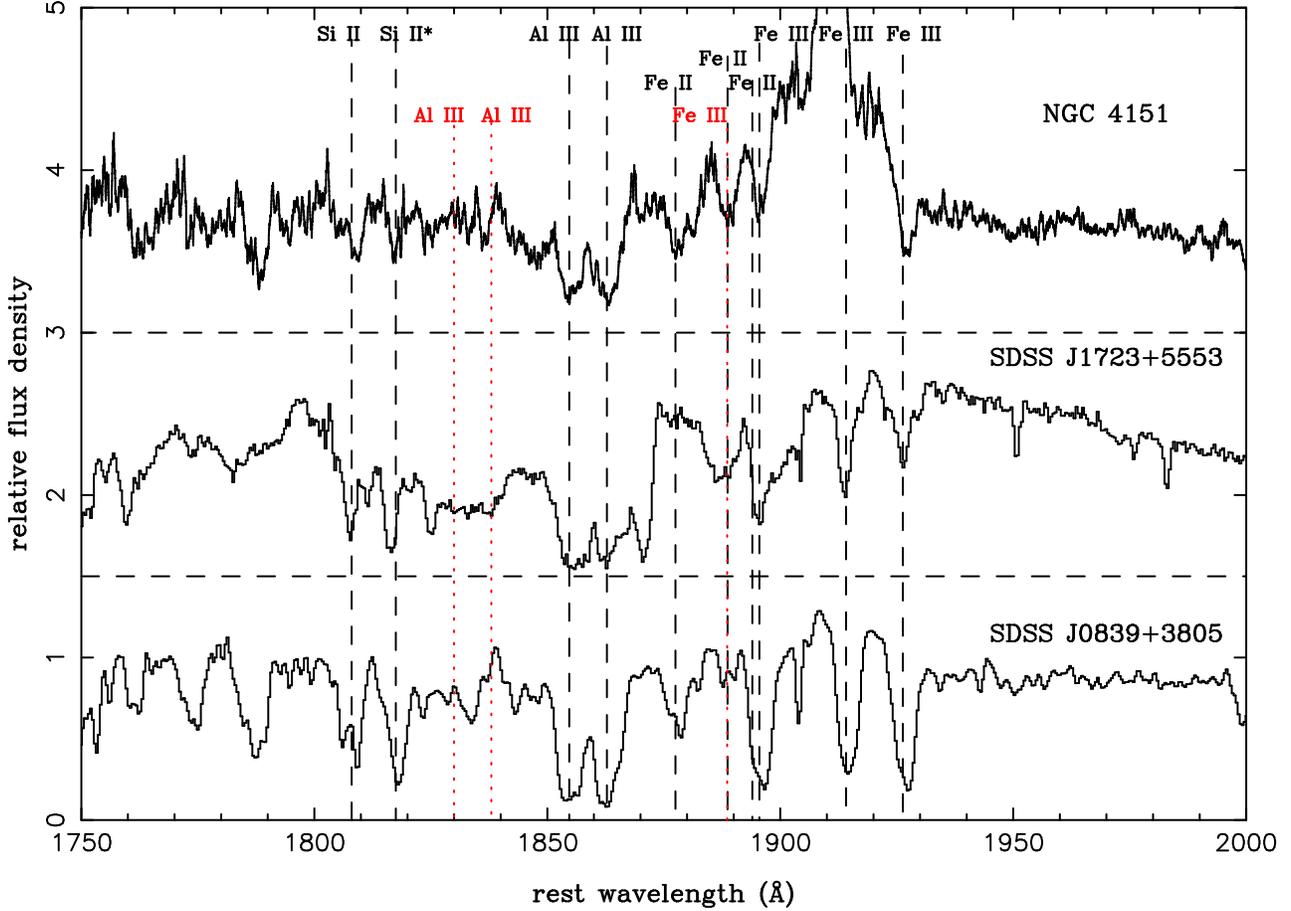}
  \end{center}
\caption{Comparison of the spectrum of BAL AGN between 1700 and 2000 {\AA}.
Every spectrum is shifted to the frame of redshift of Al \emissiontype{III} 
absorption lines, and normalized by its continuum level.
The redshift of 2.0942 is adopted for SDSS~1723+5553. 
Ordinate is a relative flux density in units of erg s$^{-1}$ cm$^{-2}$ \AA $^{-1}$. 
The spectra of SDSS~1723+5553 and NGC~4151 are offseted by +1.5 and +3.0, respectively, for clarity, 
and their zero levels are shown by horizontal dashed lines. 
The vertical dashed lines indicate the positions of Si \emissiontype{II} 1808.0, Si \emissiontype{II}* 1817.5, Al \emissiontype{III} 1854.7, Al \emissiontype{III} 1862.8,
Fe \emissiontype{II} UV125 1877.5, Fe \emissiontype{II} UV125 1888.7, Fe \emissiontype{II} UV125 1894.0, 
Fe \emissiontype{III} 1895.5, Fe \emissiontype{III} 1914.1, and Fe \emissiontype{III} 1926.3.
The $z=2.053$ system of Al \emissiontype{III} 1854.7, Al \emissiontype{III} 1862.8, and Fe \emissiontype{III} 1914.1
are indicated by vertical red dotted lines.
The $z=2.053$ system of Fe \emissiontype{III} 1914.1 coincides with Fe \emissiontype{II} UV125 1888.7 at $z=2.0942$.}
\label{fig5}
\end{figure}

The Al \emissiontype{III} $\lambda\lambda$ doublet $1854.7, 1862.8$ 
with a redshift of $2.053$ corresponds to the trough between 5660 {\AA} and 5690 {\AA} in the 
observed frame.
I compare the region between 1700 and 2000 {\AA} of SDSS~1723+5553 with
those of other Balmer BAL AGN (NGC~4151 and SDSS~J0839+3805)
\footnote{No spectrum between 1700 and 2000 {\AA} is currently available for SDSS~J1259+1213.}. 
The spectrum of NGC~4151 was taken in 1999 July when Fe \emissiontype{II} absorption 
lines were strong.
It has been published in \citet{Kra01}.
I retrieve it from the Multi-mission Archive at the Space Telescope Science 
Institute. 
The spectrum of SDSS~J0839+3805 is taken from SDSS database.
There exist no strong absorption lines at rest 1830 and 1840 {\AA}  in the spectra of NGC~4151 and SDSS~J0839+3805 (figure \ref{fig5}).
Therefore, the trough between 1830 and 1840 {\AA} in the rest frame of $z=2.0942$
must be Al \emissiontype{III} $\lambda\lambda1854.7, 1862.8$ at $z=2.053$.
\par
\begin{figure}
  \begin{center}
    \FigureFile(170mm,140mm){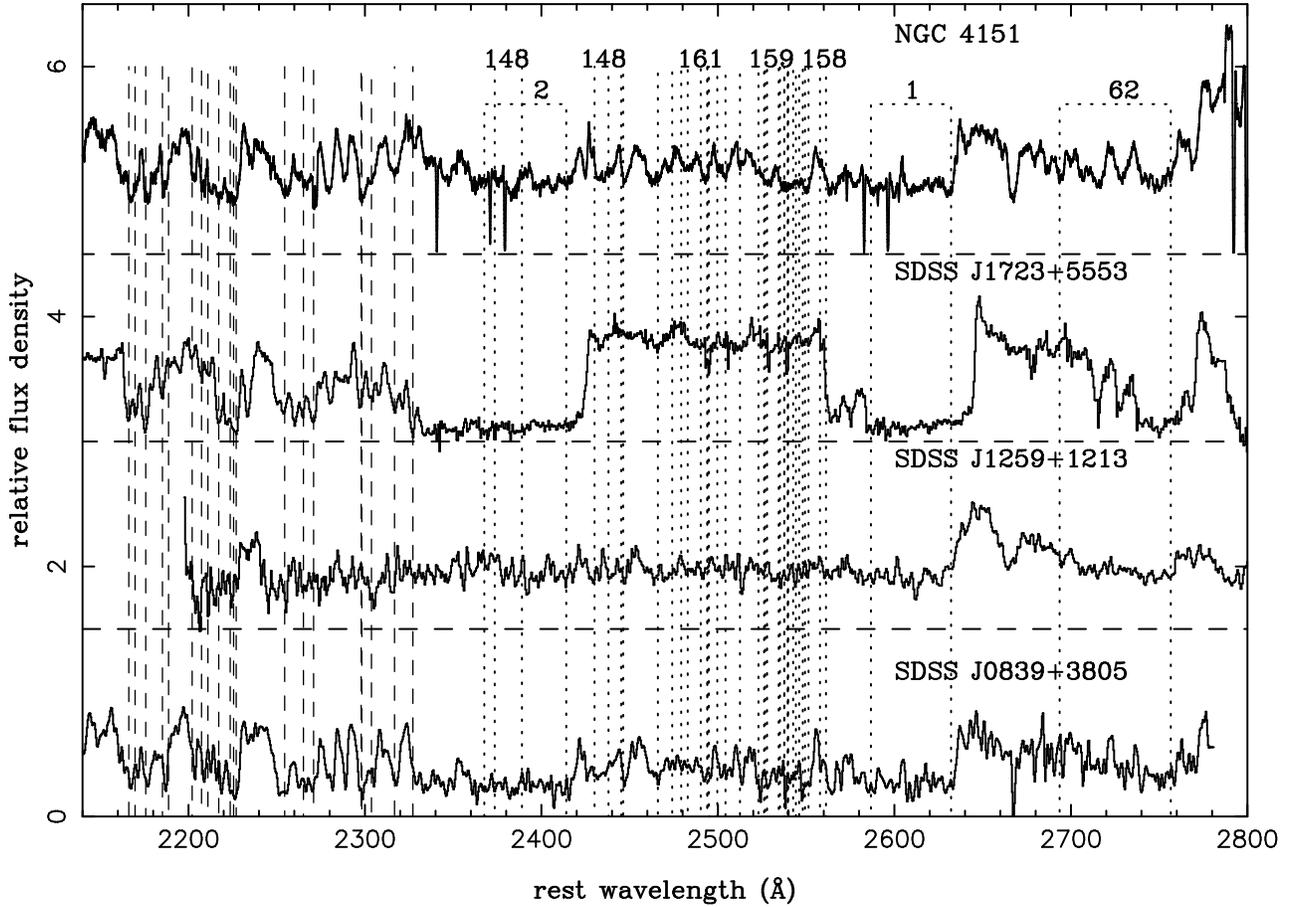}
  \end{center}
\caption{Comparison of rest near UV spectra of Balmer BAL AGN.
Every spectrum is shifted to the frame of redshift of Al \emissiontype{III} 
absorption lines, and normalized by its continuum level. 
Ordinate is a relative flux density in units of erg s$^{-1}$ cm$^{-2}$ \AA $^{-1}$.
The spectra of SDSS~1259+1213, SDSS~1723+5553 and NGC~4151 are offseted by +1.5, +3.0,  and +4.5, respectively, for clarity, 
and their zero levels are shown by horizontal dashed lines. 
The vertical dashed lines between 2180 {\AA} and 2340 {\AA}
indicate Ni~\emissiontype{II} absorption lines.
The dotted vertical lines indicate Fe~\emissiontype{II} absorption lines,
and the numbers indicate multiplets
}
\label{nearUV}
\end{figure}

The absorption at $\sim 5840$ {\AA} corresponds to Fe \emissiontype{III} UV34 $\lambda1914.1$ at $z=2.053$.
However, the wavelength of Fe \emissiontype{III} UV34 $\lambda1914.1$ at $z=2.053$
is 5843.6 {\AA}, which
is very close to the wavelength of Fe \emissiontype{II} UV125 $\lambda1888.7$
at $z=2.0942$, 5844.0 \AA.
Fe \emissiontype{II} UV125 $\lambda\lambda1877.5, 1888.7,$ and $1894.0$
is a triplet whose 
lower term ($a ^{2}H$) are 2.51--2.57 eV above ground.
The oscillator strength of Fe \emissiontype{II} UV125 is not available,
however, the Fe \emissiontype{II} UV125 $\lambda1877.5$ absorption line is stronger than 
or similar to the $\lambda1888.7$ absorption line
in the spectra of NGC~4151 and SDSS~0839+3805 (figure \ref{fig5}).
On the other hand, in the spectrum of SDSS~J1723+5553,
the Fe \emissiontype{II} UV125 $\lambda1877.5$ at $z=2.0942$ absorption is too weak to identify (figure \ref{fig5}).
Also, Fe~\emissiontype{II} UV125 $\lambda1888.7$ at $z=2.0942$ must be weak, too.
Other Fe II multiplets from excited terms ($b^{4}P$ and $a^{4}H$) of 2.6-- 2.7 eV above ground, for example, 
Fe II UV148, 158, 159, and 161 between 2430 and 2560 {\AA}
 are weak in SDSS~1723+5553 (figure \ref{nearUV}).
In the other three Balmer BAL AGNs those absorption lines are strong and 
most of the continuum light is absorbed. 
Thus, the absorber of SDSS~1723+5553 has a lower density and/or lower temperature than the other three Balmer BAL AGNs. 
I can thus ignore the Fe \emissiontype{II} UV125 $\lambda1888.7$ absorption at $z=2.0942$.
The absorption at $\sim 5840$ {\AA} must be Fe \emissiontype{III} UV34 $\lambda1914.1$ absorption at $z=2.053$.
\par
\begin{figure}
  \begin{center}
    \FigureFile(135mm,110mm){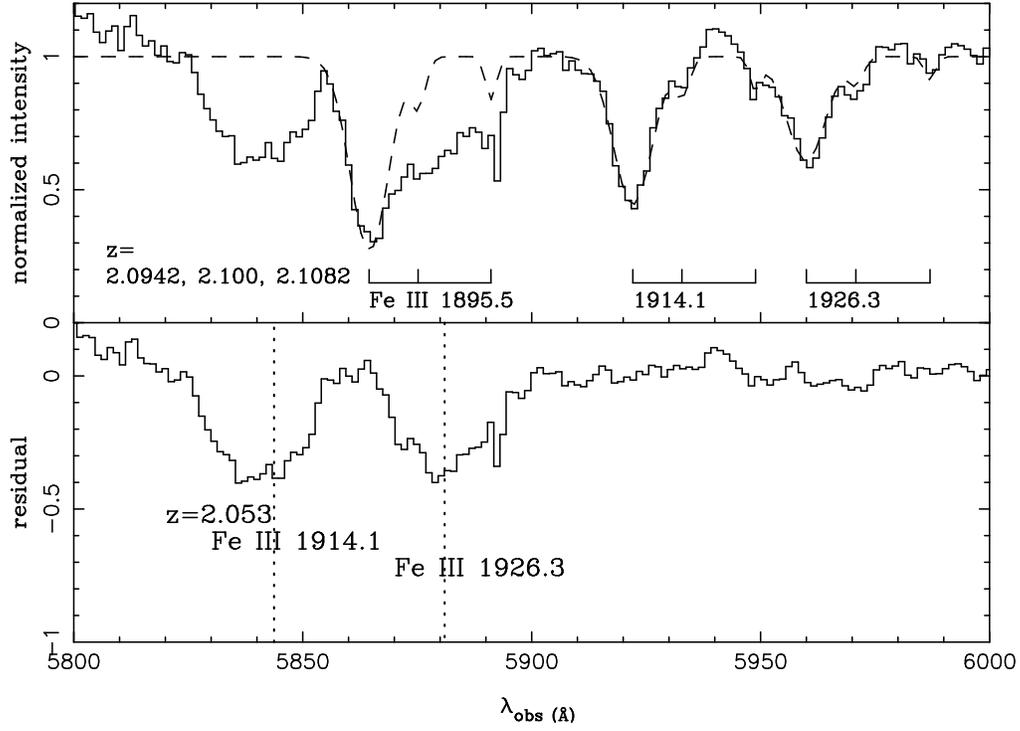}
  \end{center}
  \caption{Fe \emissiontype{III} absorption at $z=2.053$.
Top: The normalized spectrum of SDSS~J1723+5553 between 5800 {\AA} and 6000 {\AA} 
in observed frame is shown in the histogram.
The best fit of Fe \emissiontype{III} absorption at $z=2.0942, 2.100$ and 2.1082
is shown in the dashed line.
Note that the region between 5800 {\AA} and 5900 {\AA} was not used for the fitting.
Bottom: The residual spectrum after subtraction of Fe \emissiontype{III} 
absorption at $z=2.0942, 2.100$ and 2.1082 is shown in the histogram.
The expected positions of Fe \emissiontype{III} $\lambda\lambda 1914.1, 1926.3$
absorption lines at $z=2.053$
are indicated by the dotted lines.
They clearly correspond to two absorption lines at $z=2.053$.}
\label{FeIIImodel}
\end{figure}

The wavelength of Fe \emissiontype{III} UV34 $\lambda1926.3$ at $z=2.053$
is between Fe \emissiontype{III} UV34 $\lambda1895.5$ at $z=2.100$ and 2.1082.
I model Fe \emissiontype{III} UV34 $\lambda1895.5$ {\AA} absorption
at $z=2.0942, 2.100$ and 2.1082, and subtract it from the spectrum.
First, I normalize the spectrum of SDSS~1723+5553 between 5800 and 6000 {\AA} with a Legendre polynomial.
The normalized spectrum of SDSS~1723+5553 is shown in the upper panel of
figure \ref{FeIIImodel}.
In order to make a model profile of Fe \emissiontype{III} UV34 
$\lambda1895.5$ {\AA} absorption at $z=2.0942, 2.100$ and 2.1082, 
I fit profiles of Fe \emissiontype{III} UV34 
$\lambda\lambda1914.1,1926.3$ at $z=2.0942, 2.100$ and 2.1082.
I fix the ratio of intensity of the absorption as the ratio of each 
oscillator strength of the transition.
The widths of the absorption troughs are same among components of the same 
redshift. 
The redshifts are fixed among the three transitions.
I thus fit the normalized spectrum
between 5900 {\AA} and 6000 {\AA} with a continuum and two absorption lines
(Fe \emissiontype{III} UV34 $\lambda\lambda1914.1,1926.3$).
I model Fe \emissiontype{III} UV34 $\lambda1895.5$ based on the fitting
results of Fe \emissiontype{III} $\lambda\lambda1914.1,1926.3$.
The model profile of Fe \emissiontype{III} UV34 
at $z=2.0942, 2.100$ and 2.1082
is shown as the dashed line in the upper panel of
figure \ref{FeIIImodel}.
The fit is good for Fe \emissiontype{III} UV34 
$\lambda\lambda1914.1, 1926.3$.
The result of the subtraction is shown in the lower panel of 
figure \ref{FeIIImodel}.   
The two absorption features remain at the wavelength of 
Fe \emissiontype{III} UV34 
$\lambda\lambda1914.1, 1926.3$ at $z=2.053$.
Fe \emissiontype{III} UV34 $\lambda1895.5$ at $z=2.053$ unfortunately
overlaps with Al \emissiontype{III} emission and the absorption line 
(figure~\ref{fig4}).
It is difficult to deblend them.
\citet{Hal02} has pointed out unidentified absorption lines at 5840 and 5680 {\AA} (observed frame) in SDSS~J1723+5553 spectrum.
Those unidentified absorption lines I identify as the $z=2.053$ system of 
Fe~\emissiontype{III} $\lambda 1914.1$
and Al~\emissiontype{III} $\lambda\lambda1854.7, 1862.8$, respectively.
\par
I fit Fe~\emissiontype{III} $\lambda 1914.1, 1926.3$
and Al~\emissiontype{III} $\lambda\lambda1854.7, 1862.8$ at $z=2.053$ with gaussians
in order to measure their widths.
The Fe~\emissiontype{III} absorption lines are fitted with two gaussians of the same
 width and redshift.
The Al~\emissiontype{III} absorption lines are also fitted with the same way.
The FWHM$_{\rm true}$ of Fe~\emissiontype{III} and Al~\emissiontype{III} are 
$1030 \pm 60$ and $1350 \pm 90$ km s$^{-1}$, with
redshifts of $2.0522 \pm 0.0002$, $2.0517 \pm 0.0004$, respectively.
The FWHMs are consistent with those of the Balmer absorption lines.

\section{Conclusion}
I present the discovery of Balmer-series absorption lines from H$\alpha$ to H9 in SDSS~J1723+5553
by near-infrared spectroscopy.
The Balmer-line absorption is at $z=2.053$ and blueshifted by 5370 km s$^{-1}$ from
the Balmer emission lines.
I derive the column density of neutral hydrogen of $5.2\times10^{17}$ cm$^{-2}$
by the curve of growth method. 
The Balmer-line absorption is more than $4000$ km s$^{-1}$
blueshifted from the previously known UV absorption lines.
I search for the same velocity component seen in the UV absorption lines with Balmer-line absorption.
I find the component at $z=2.053$ in Al \emissiontype{III} and Fe \emissiontype{III} absorption lines.
\par
High resolution near-infrared spectroscopy is needed to
estimate a more precise column density of neutral hydrogen and 
covering factor of the absorber.
If we can deblend He \emissiontype{I} $\lambda3890$ from H8 absorption,
we will be able to estimate ionized helium column density.
He \emissiontype{I} $\lambda3890$ absorption arises from the metastable $2~^{3}S$ 
level of He$^{0}$.
The population of the metastable $2~^{3}S$ is balanced by recombination to all triplet levels of He$^{+}$
and collisional transition to other levels. 
Also, search for other UV absorption at $z=2.053$  
by high resolution optical spectroscopy would be interesting.
Farther if we can identify absorption troughs from the excited metastable level of Fe \emissiontype{II} and/or Si \emissiontype{II},
the derived relative populations will allow us to estimate the electron number density
of the absorber associated with Balmer-line absorption.
When combined with photoionization modeling, the distance of the absorber from the nucleus can be estimated.
The Balmer absorption troughs translates to a large number of n=2 level hydrogen atoms,
which is result of Ly$\alpha$ trapping.
A high density and a high column density are necessary for Ly$\alpha$ trapping. 
This indicates that the outflow associated with Balmer absorption lines probably traces gas closer to nuclei.
Combination of such high velocity of 5370 km s$^{-1}$ and high column density, the mass flux of flow in SDSS~1723+5553 is probably as large as already measured in other FeLoBALs
\citep{Dunn10}. 

\bigskip
I am grateful to the staffs of Subaru Telescope for their assistance during our observations.
I also thank Toshihiro Kawaguchi and the anonymous referee for his or her helpful comments.
I would also like to thank Jay P. Dunn for patient English proof reading.
Funding for the Sloan Digital Sky Survey (SDSS) and SDSS-II has been provided by the Alfred P. Sloan Foundation, the Participating Institutions, the National Science Foundation, the U.S. Department of Energy, the National Aeronautics and Space Administration, the Japanese Monbukagakusho, and the Max Planck Society, and the Higher Education Funding Council for England. The SDSS Web site is http://www.sdss.org/.
The SDSS is managed by the Astrophysical Research Consortium (ARC) for the Participating Institutions. The Participating Institutions are the American Museum of Natural History, Astrophysical Institute Potsdam, University of Basel, University of Cambridge, Case Western Reserve University, The University of Chicago, Drexel University, Fermilab, the Institute for Advanced Study, the Japan Participation Group, The Johns Hopkins University, the Joint Institute for Nuclear Astrophysics, the Kavli Institute for Particle Astrophysics and Cosmology, the Korean Scientist Group, the Chinese Academy of Sciences (LAMOST), Los Alamos National Laboratory, the Max-Planck-Institute for Astronomy (MPIA), the Max-Planck-Institute for Astrophysics (MPA), New Mexico State University, Ohio State University, University of Pittsburgh, University of Portsmouth, Princeton University, the United States Naval Observatory, and the University of Washington.
This publication makes use of data products from the Two Micron All Sky Survey, 
which is a joint project of the University of Massachusetts and the Infrared Processing and Analysis Center/California Institute of Technology, funded by the National Aeronautics and Space Administration and the National Science Foundation.
This research has also made use of the Atomic Line List, version 2.04, available at http://www.pa.uky.edu/$\sim$peter/atomic/.

\clearpage

\clearpage 

\clearpage 

\clearpage 

\clearpage 

\clearpage 

\clearpage 

\clearpage 


\begin{table}
  \caption{Properties of Emission and Absorption Lines.}\label{tbl1}
  \begin{center}
\begin{tabular}{cccc}
\hline
     &                      &      & $\rm{FWHM}_{true}$\\
Line & Absorption/Emission & $z$ & (km s$^{-1}$)\\
\hline
H$\alpha$   & Emission & $2.1081\pm0.0003$ & $2780\pm130$ \\
H$\beta$    & Emission & $2.1080\pm0.0008$ & $3780\pm150$ \\
H$\gamma$\footnotemark[$*$]   & Emission & $2.1107\pm0.0012$ & -- \\
H$\delta$   & Emission & $2.1095\pm0.0041$ & -- \\
H$\epsilon$ & Emission & $2.0973\pm0.0057$ & -- \\
 \rm{[O \emissiontype{III}]} 5008 & Emission & $2.1095\pm0.0004$ & $1550\pm70$ \\
\hline
H$\alpha$\footnotemark[$\dagger$] & Absorption & $2.0500\pm0.0007$ & $760\pm300$\\
H$\beta$    & Absorption & $2.0527\pm0.0005$ & $450\pm130$ \\
H$\gamma$   & Absorption & $2.0530\pm0.0006$ & $< 890$ \\
H$\delta$   & Absorption & $2.0532\pm0.0008$ & $520\pm200$ \\
H$\epsilon$ & Absorption & $2.0533\pm0.0013$ & $1000\pm430$ \\
H8\footnotemark[$\ddagger$] & Absorption & $2.0517\pm0.0006$ & $< 710$ \\
H9         & Absorption & $2.0540\pm0.0015$ & $< 710$ \\
\hline
\multicolumn{4}{@{}l@{}}{\hbox to 0pt{\parbox{85mm}{\footnotesize
\par\noindent
\footnotemark[$*$] H$\gamma$ emission is affected by the atmospheric absorption lines.
\par\noindent
\footnotemark[$\dagger$] H$\alpha$ absorption is affected by the atmospheric absorption lines.
\par\noindent
\footnotemark[$\ddagger$] H8 absorption is probably blended with He \emissiontype{I} 3890 absorption line.
}\hss}}
\end{tabular}
\end{center}
\end{table}
\clearpage

\begin{table}
  \caption{Equivalent widths of Balmer absorption lines.}\label{tbl2}
  \begin{center}
\begin{tabular}{cccc}
\hline
            & $\lambda$ &     $f$                & EW$_{rest}$ \\
            & \AA       & oscillator strength & \AA         \\
\hline
H$\alpha$   & 6564.610  & 6.40E-01 & $6.4\pm1.4$\footnotemark[$*$] \\
H$\beta$    & 4862.683  & 1.19E-01 & $5.3\pm0.53$ \\
H$\gamma$   & 4341.684  & 4.46E-02 & $2.4\pm0.31$ \\
H$\delta$   & 4102.892  & 2.21E-02 & $3.0\pm0.38$ \\
H$\epsilon$ & 3971.195  & 1.27E-02 & $1.8\pm0.28$ \\
H8          & 3890.151  & 8.03E-03 & $3.2\pm0.33$\footnotemark[$\dagger$] \\
H9          & 3836.472  & 5.43E-03 & $0.76\pm0.27$ \\
\hline
\multicolumn{4}{@{}l@{}}{\hbox to 0pt{\parbox{85mm}{\footnotesize
\par\noindent
\footnotemark[$*$] H$\alpha$ absorption is affected by the atmospheric absorption lines.
\par\noindent
\footnotemark[$\dagger$] H8 absorption is probably blended with He \emissiontype{I} 3890 absorption line.
}\hss}}
\end{tabular}
\end{center}
\end{table}
\clearpage

\end{document}